\newcommand{\bld}[1]{\mbox{\boldmath{$#1$}}}

\newcommand{\kalb}{{\textstyle {1\over 2}}}
\newcommand{\beg}{\begin{equation}\label}
\newcommand{\en}{\end{equation}}

\newcommand{\mun}{\mu_{nr}}
\newcommand{\nds}{{n_D^*}}
\documentstyle[preprint,aps,12pt,floats]{revtex}
\begin{document}
\draft
\tighten

\title{$\cal CPT$-invariant two-fermion Dirac equation with extended hyperfine 
operator}
\author{Viktor Hund and Hartmut Pilkuhn}

\address{Institut f\"ur Theoretische Teilchenphysik, Universit\"at,
D-76128 Karlsruhe, Germany\\ e-mails: vh \& hp@particle.physik.uni-karlsruhe.de}

\maketitle

\begin{abstract}{For the S-states of muonium and positronium, the hyperfine
shifts to the order $\alpha^6$ of a recently derived two-fermion equation with 
explicit $\cal CPT$-invariance are checked against the results of a 
nonrelativistic reduction, and the leading $\alpha^8$ shifts are calculated. 
An additional hyperfine operator is discovered which can milden the 
singularity for $r\to 0$ of the Dirac hyperfine operator, such that the 
resulting extended operator can be used nonperturbatively. The binding 
correction to magnetic moments is mentioned.
\begin{center}{PACS number: 03.65.Pm, 11.30 Er, 32.60.+i}\end{center}
}\end{abstract}
\vspace{0.8cm}

\begin{section}{Introduction}\label{ch1}
The energy levels of one-electron atoms are well described by the Dirac 
equation, particularly when the nuclear hyperfine operators are included,
and when radiative corrections are computed from the vector potential
operator that is part of the electron momentum operator (Erickson 1977).
But the Dirac equation applies also to muonium $(e^-\mu^+)$, where the ``nucleus`` is a structureless muon. A simple recoil correction reduces the electron
mass $m_1$ by a factor $m_2/m=m_2/(m_1+m_2)$, where $m_2$ is the muon mass.
The resulting reduced mass $\mu_{nr}=m_1m_2/m$ is familiar from 
the two-body Schr\"odinger equation, but apart from the limit $m_2/m\approx
1-m_1/m_2$ (Braun 1973), the correctness of this factor in the Dirac equation
has not been proven.

For positronium $(e^-e^+)$, $m_2/m=1/2$ excludes a perturbative treatment of
recoil. In the past, precise calculations of muonium and positronium have
used the Bethe-Salpeter equation, which treats both particles relativistically
(Sapirstein and Yennie 1990). However, the complicated higher-order radiative
corrections contain only few relativistic effects. An alternative method has
been elaborated in which all relativistic effects are treated perturbatively,
by effective operators added to the nonrelativistic two-body Schr\"odinger
Hamiltonian $\bld{p}^2/2\mu_{nr}+V(r)$ (Caswell and Lepage 1986). By this 
``nonrelativistic quantum electrodynamics method'' (NRQED), the energy levels 
of muonium
and positronium have been calculated to the order $\alpha^6$ (Pachucki 1997).
To that order at least, all results of the Dirac equation with the above 
reduction factor are confirmed at the lowest nonvanishing power of 
$\mu_{nr}/m$, 
not only in the coarse and fine structures, but also in the hyperfine 
structure. In the calculation of the muonium hyperfine splitting, the term 
$\alpha^8\mu_{nr}^2/m$ must also be included, but only in the approximation 
$\mu_{nr}^2/m\approx m_1^2/m_2$ (Kinoshita 1998). It is clearly more convenient
to rely on the correctness of the Dirac expectation value of the relativistic
hyperfine operator than to extract this term from four-photon exchange
graphs in NRQED.

Recently, a relativistic two-body equation with only eight components and 
explicit $\cal CPT$-invariance has been
constructed from the QED scattering matrix (H\"ackl et al 1998) for two 
structureless fermions (a short rederivation is presented in appendix A).
 Although both particles are treated symmetrically, 
the equation looks like a one-body Dirac equation, in which the mass $m_1$
and energy $E_1$ of particle 1 are replaced by a relativistic reduced mass
$\mu$ and reduced energy $\varepsilon$:
\beg{1}\mu=m_1m_2/E,\quad\varepsilon/\mu=(E^2-m_1^2-m_2^2)/2m_1m_2,\en
where $E$ is the total cms energy $(c=1)$. This equation has predictive power for all
values of $\mu_{nr}/m$, including the positronium value $\mu_{nr}/m=1/4$.
For the nS-states, $E$ is $\mu_{nr}$ times a polynomial in $\mu_{nr}/m$, which
has the order 2 to the order $\alpha^6$ and 3 to the order $\alpha^8$. In this
paper, the second-order polynomial is checked against 
the NRQED results, and the third-order polynomial is constructed to an extent 
that may become relevant in the near future.

For a simple understanding of the origin of the polynomial, the hyperfine operator is neglected in section 2. The energy levels are then obtained
 from those of the static Dirac equation, $E_1/m_1=f(n^*)$, by the replacement
$E_1/m_1\to\varepsilon/\mu$, and by the subsequent evaluation of $E-m$ as a 
power series in $f-1$. The resulting terms of order $\mu_{nr}$ and
$\mu_{nr}^2/m$ give directly the hyperfine-averaged energy levels, while the 
order $\mu_{nr}^3/m^2$ contains additional second-order hyperfine effects in
 the hyperfine-averaged levels, beginning at the order $\alpha^6$.

The argument $n^*$ in $f(n^*)$ denotes the effective principal quantum number,
$n^*=n+\delta \ell$, where $\delta \ell$ is a (negative) quantum defect. With
$\delta \ell$ unspecified, the results of section 2 are more general than 
their derivation. This is so because the function 
\beg{2}f(n^*)=(1+\alpha^2/n^{*2})^{-1/2}\approx 1-\alpha^2/2n^{*2}\en
solves not only the Dirac equation, but also the Klein-Gordon equation. The 
latter case has $\delta \ell_{KG}=[(\ell+1/2)^2-\alpha^2]^{1/2}-\ell-1/2$, 
while the Dirac case has the orbital angular momentum $\ell$ replaced by 
$j=1/2$ for $\ell=0$ and by $j=\ell\pm 1/2$ for $\ell>0$:
\beg{3}\delta \ell_D= \gamma-j-\kalb,\quad\gamma=[(j+\kalb)^2-\alpha^2]^{1/2}.
\en
Thus expressions (1) and (2) occur also in the bound states of two 
spinless particles
such as $\pi^-K^+$ and $\pi^-\pi^+$. On the other hand, bound states of one
lepton and one spinless particle ($\mu^-\pi^+$ or $e^-$ $ ^4$He) require a more
complicated equation with an asymmetric dependence on $m_1$ and $m_2$, which
produces the socalled Barker-Glover term at the order $\alpha^4$.(Atomic
hydrogen requires also an asymmetric interaction, due to the proton's
large anomalous magnetic moment.) In section 3, the expectation values of 
the Dirac
 hyperfine operator of the two-fermion Dirac equation are expanded in powers of
$f(n_D^*)-1$ to order $\alpha^8\mu_{nr}^3/m^2.$ The first-order expansion for
Zeeman operators is mentioned, in which the $\cal CPT$ origin is particularly 
evident.

Radiative corrections are ignored so far, but for the more complicated graphs,
they may again be evaluated nonrelativistically. The simpler graphs can be
evaluated by Dirac methods also for the new equation. There are however other
effective two-photon exchange operators that must be added in the two-fermion
equation. For $\ell>0$, they are $7\alpha^2/6\pi\mu_{nr}mr^3$ and $\alpha^2L^2
/2\mu_{nr}^2mr^4$. The first of these contributes at the order $\alpha^5/\pi$,
where it is part of the socalled Salpeter shift. Its precise form for S-states
is $7\alpha^2p_G^2/6\pi\mu_{nr}$ (Gupta et al 1989),
\beg{4}mp_G^2=-\nabla^2\left(\frac{\ln(\gamma\mu r)}{r}\right)
=-\left[\partial_r,\left[\partial_r,
\frac{\ln(\gamma\mu r)}{r}\right]\right]-\frac{2}{r}\left[\partial_r,
\frac{\ln(\gamma\mu r)}{r}\right],\en
with $\gamma=e^C$ and $C=$ Euler's constant 0.577...,with the prescription that
expectation values of $p_G^2$ are calculated by partial integration.
This operator is mentioned here because its relativistic analogue occurs in
the hyperfine interaction at the order $\alpha^6\mu_{nr}^3/m^2$, to be 
discussed in section 4.
\end{section}

\begin{section}{The energy levels of the two-fermion Dirac-Coulomb equation}
\label{ch2}
When the hyperfine operators are neglected, the resulting Dirac-Coulomb 
equation has exact solutions. In units of the relativistic reduced mass $\mu$
(\ref{1}), the equation reads $(\hbar=c=1)$
\beg{5} \left(\frac{\varepsilon}{\mu}-V(\rho)-\beta-\gamma_5\bld{\sigma}_1
{\bld p}_\rho\right)\psi_{DC}=0,\quad {\bld\rho}=\mu{\bld r},\quad V=-
\frac{\alpha}{\rho} \en
with $\beta=\gamma^0,\;\; \gamma_5\bld{\sigma}_1=\bld{\alpha}=\gamma^0
\bld{\gamma}$ as usual, and ${\bld p}_\rho={\bld p}/\mu$.\\
The solutions are
\beg{6}\frac{\varepsilon}{\mu}=f_D=\left(1+\frac{\alpha^2}{{n_D^*}^2}\right)^{
-1/2},\quad n_D^*=n+\gamma-j-\frac{1}{2}\en
with $\gamma$ defined in (\ref{3}). With the definitions (\ref{1}) 
of $\mu$ and 
$\varepsilon$, the binding energy $E-m$ is expanded in powers of $f_D-1$ as 
follows:
\begin{eqnarray}
E-m &=& \sqrt{m^2+2m_1m_2(f_D-1)}-m=\mun(f_D-1)\nonumber\\
&&- \frac{1}{2}(f_D-1)^2\,\frac{\mun^2}{m}+\frac{1}{2}(f_D-1)^3\,
\frac{\mun^3}{m^2} \dots\label{7}\end{eqnarray}
The precise expansion parameter is $m_1m_2(\alpha/n_D^*m)^2$, but as $n_D^*$ is
also expanded in terms of $\alpha$ at a later stage, the mass dependence must 
be kept explicitly in (\ref{7}). To the order $\alpha^8$, one needs
\begin{eqnarray} 
f_D-1\; &=& -\frac{\alpha^2}{2\nds^2}\left[1-\frac{3}{4}\frac{\alpha^2}{\nds^2}
+\frac{5}{8}\frac{\alpha^4}{\nds^4}\left(1-\frac{7}{8}\frac{\alpha^2}{\nds^2}
\right)\right] \label{8}\\
(f_D-1)^2 &=& \frac{\alpha^4}{4\nds^4}\left[1-\frac{3}{2}\frac{\alpha^2}{
\nds^2}+\frac{29}{16}\frac{\alpha^4}{\nds^4}\right] \label{9}\\
(f_D-1)^3 &=& -\frac{\alpha^6}{8\nds^6}\left[1-\frac{9}{4}\frac{\alpha^2}{
\nds^2}\right],\quad  (f_D-1)^4= \frac{\alpha^8}{16\nds^8}\label{10}
\end{eqnarray}
It has been pointed out that the simple relation (\ref{1}) between $\varepsilon
/\mu$ and $E^2$ makes the expansion of $E^2-m^2$ simpler than that of $E-m$, 
but for the check against present NRQED it is necessary to expand $E-m$. In 
particular, one sees that for $\alpha^6$ the mass dependence ends at $\mun^3/
m^2$. This order in $\mun/m$ receives contributions also from second-order 
hyperfine effects, which are discussed in the next section. \\
In the frame of the two-fermion Dirac equation, all terms discussed so far 
originate from the one-photon exchange Born graph. Additional two-photon 
exchange graphs contribute already at the order $\alpha^5/\pi$. These have only
been evaluated with nonrelativistic approximations, where most of them result 
in the operator $\delta(\bld{r})$, with matrix elements proportional to 
$\delta_{\ell 0}/n^3$. The only two-photon exchange operator with a more 
complicated $n$-dependence is the ``Gupta operator'' (\ref{4}), which has the 
following expectation value for $\ell = 0$
\beg{11}\langle p_G^2\rangle_{\ell=0}=2\mun^3\frac{\alpha^3}{mn^3}\left[
\ln{\frac{2\alpha}{n}}+{\sum_{i=1}^n}\frac{1}{i}-\frac{1}{2n}+
\frac{1}{2}\right]=\frac{2}{m}\left(\frac{\mun\alpha}{n}\right)^3\left[
\ln{\frac{2\alpha}{n}}+\Psi+C+\frac{n+1}{2n}\right]\en
with $\Psi=\Psi(n)=d \Gamma(n)/dn$. At this order in $\alpha$, there is no 
finite-range two-photon exchange operator in the hyperfine structure. At the 
next order $\alpha^6$, there is no finite-range three-photon exchange operator 
for S-states
in the Dirac-Coulomb equation, but there is one such operator to be added to 
the Dirac hyperfine operator, which is discussed in section 4. \\
All terms in (\ref{7}) that require only the nonrelativistic values, $f_D-1=-
\alpha^2/2n^2,\;\;(f_D-1)^2=\alpha^4/4n^4$ etc. are valid for all combinations
of spins and magnetic moments. At the order $\alpha^4$, there is only one such 
term, originating from the second term in the expansion (\ref{7}). It was 
originally  discovered by Bechert and Meixner (1935). The complete series 
(\ref{7}) was discovered for two spinless particles by Brezin et al (1970), 
with $\nds$ replaced by the appropriate Klein-Gordon value $n_{KG}^*$. 
It could not yet be confirmed experimentally. For positronium, on the other
hand, the 
agreement between the NRQED calculation to the order $\alpha^6$ and experiment 
is not perfect, but it seems fair to say that all terms in the expansion 
(\ref{7}) are now checked, for the extreme case $\mun/m=1/4$ (Pachucki and 
Karshenboim 1998, Czarnecki et al 1999).
\end{section}

\begin{section}{Dirac hyperfine splitting}\label{ch3}
Inclusion of the hyperfine operator in the dimensionless reduced Dirac equation
(\ref{5}) leads to (H\"ackl et al 1998)
\beg{12} \left[\frac{\varepsilon}{\mu}-V(\rho)-\beta-\gamma_5(\bld{\sigma}_1
+\bld{\sigma}_{hf}^{(1)}){\bld p}_\rho\right]\psi=0,\quad 
\bld{\sigma}_{hf}^{(1)}=-i\bld{\sigma}_1\times\bld{\sigma}_2 V\frac{\mu}{E}\en
By standard perturbation theory, the first-order shift of $\varepsilon/\mu$ 
caused by the hyperfine operator is (Rose 1961)
\beg{13}\left(\frac{\varepsilon}{\mu}\right)_{hf}^{(1)}= \alpha^4\frac{\mu}{E}
f^{(1)}_{hf,D}, \quad f^{(1)}_{hf,D}=\frac{4(f-j)}{f+1/2}f_D^3\frac{(j+1/2)^2 
f_D-\kappa_D/2}{\gamma(2\gamma+1)(2\gamma-1)}\en
where $f=j\pm 1/2$ is the total angular momentum, and $\kappa_D=2(\ell-j)
(j+1/2)$.
Any small shift $\delta(\varepsilon/\mu)=\delta(E^2/2m_1m_2)$ corresponds
to a small shift $\delta E$,
\beg{13a}\delta(E^2)=2E\delta E.\en
Thus the first-order hyperfine shift is
\beg{14}E^{(1)}_{hf}=\mu\left(\frac{\varepsilon}{\mu}\right)_{hf}^{(1)}\approx
2\frac{\mun^2}{m}\alpha^4 f^{(1)}_{hf,D}\left[1-3\frac{\mun}{m}(f_D-1)+
\frac{15}{2}\left(\frac{\mun}{m}\right)^2(f_D-1)^2\right]\en
In the hyperfine structure of muonium, $\mun/m<\alpha$ makes the last term in 
(\ref{14}) negligible, while $f_D-1$ is required at most to the order 
$\alpha^4$. Insertion of $\nds^{-2}\approx n^{-2}+\alpha^2/n^3(j+1/2)$ gives
\beg{15}E^{(1)}_{hf}=2\frac{\mun^2}{m}\alpha^4 f^{(1)}_{hf,D}\left[1+3
\frac{\mun}{m}\frac{\alpha^2}{2n^2}\left(1-\frac{3}{4}\frac{\alpha^2}{n^2}+
\frac{\alpha^2}{n(j+1/2)}\right)\right].\en
Turning now to $f^{(1)}_{hf,D}$, we find to order $\alpha^6$ and for $j=\ell+
\kalb$
\begin{eqnarray} 
f^{(1)}_{hf,D} &=& \frac{2(f-j)}{(2f+1)n^3j(j+1/2)}\;(1+\alpha^2 c_2 +\alpha^4 
c_4), \label{16}\\
&&c_2=\frac{1}{j(j+1)}+\frac{1}{2(j+1/2)^2}+\frac{3}{2n(j+1/2)}-\frac{3}{2n^2}
-\frac{j+1/2}{2n^2(j+1)}. \label{17}
\end{eqnarray}
For $j=\frac{1}{2}$ (nS-states), the $\alpha^4$ correction in (\ref{16}) 
becomes
\beg{18}c_4=\frac{1}{4}\left(\frac{203}{18}+\frac{25}{2n}-\frac{67}{9n^2}-
\frac{55}{3n^3}+\frac{21}{2n^4}\right).\en
For the ``circular'' states with $n=j+\frac{1}{2}$ (which include the ground 
state) (\ref{6}) gives $\nds=\gamma,\;\; f_D=\gamma/n$, in which case 
$f^{(1)}_{hf,D}$ is greatly simplified:
\beg{19}f^{(1)}_{hf,D}\left(j=n-\frac{1}{2}\right)=4\frac{f-j}{f+1/2}\,
\frac{n\gamma+n/2}{n^3\gamma(2\gamma+1)(2\gamma-1)}=2\frac{f-j}{f+1/2}\,
\frac{1}{n^2(2\gamma^2-\gamma)}.\en
This expression was used in the calculation of the muonium hyperfine splitting.
For $n=1$, the $\alpha^4$-component of the last bracket is $17\alpha^4/4$, in agreement with (\ref{18}). It amounts to 12 ppb in the muonium hyperfine 
splitting, $\Delta\nu=$ 4463 302 617(510) Hz (the error of 510 Hz arises from 
the uncertainty of $m_\mu$). As both $c_2(j=\kalb)$ and its recoil correction
(\ref{14}) are positive, inclusion of the latter one gives an (insignificant)
increase of $\Delta\nu$. A recent experimental determination
 (Liu et al 1999) gives $\Delta \nu=$ 4463 302 776(51) Hz.

The hyperfine splitting in P-states with total angular momentum $f=1$ is 
complicated by the mixing of $j=\frac{1}{2}$ and $j=\frac{3}{2}$ states at the 
order $\alpha^4$; in this case the $\mun/m$-expansion for a given power of 
$\alpha$ does not terminate (Pilkuhn 1995). Fortunately, the S-states are 
simpler. At the 
order $\alpha^6$, the n$^3S_1$ states have a mixing between $j=\frac{1}{2}$ 
and $j=\frac{3}{2}$ (the S-D-mixing), and all S-states have contributions from 
the squares of the diagonal hyperfine matrix elements. Both effects are of 
second order in the hyperfine operator; their energy shifts 
$(\varepsilon/\mu)^{(2)}_{hf}$ may be calculated with Schr\"odinger wave 
functions and then simply added to the higher-order relativistic terms of 
$(\varepsilon/\mu)^{(1)}_{hf}$.
The S-D-mixing contributes $-4\alpha^6\mun^3/9m^2n^5$ to the hyperfine 
splitting, which combines with $+4\alpha^6\mun^3/m^2n^5$ from the second term 
of (\ref{15}) into a total of  $32\alpha^6\mun^3/9m^2n^5$, in agreement with 
the NRQED result (Pachucki 1997). This is presently the only confirmation of the 
expansion (\ref{14}). The squares of the diagonal matrix elements may be 
calculated from an effective Schr\"odinger equation, which in units of $\mu$ is
\beg{20}({\bld{\tilde p}^2_\rho}/2+V-\varepsilon/\mu+1)\psi_{Sch}=0\en
where ${\bld{\tilde p}}^2_\rho$ comprises $\bld{p}^2_\rho$ and all other 
interaction that may be approximated by a $\rho^{-2}$-potential:
\beg{21}{\bld{\tilde p}}^2_\rho=-(\partial_\rho+1/\rho)^2+
\ell'(\ell'+1)/\rho^2\en
The case at hand has
\beg{22}\ell'-\ell\equiv \delta\ell=\alpha^2\left(-\frac{1}{2j+1}+
\frac{2(f-j)}{f+1/2}\,\frac{\mun}{m}\right)\equiv\delta \ell_{j}+\delta
\ell_{hf}\en
To order $\alpha^6$, one obtains
\beg{23} E-m=-\alpha^2(1-2\delta \ell/n+3\delta \ell^2/n^2)\mu_{nr}/2n^2
-\alpha^4\mu_{nr}^2 (1-4\delta \ell/n)/8mn^4-\alpha^6\mu_{nr}^3/16m^2n^6,\en
where the term $-3\alpha^2\delta\ell^2\mun/2n^4$ contains the desired 
$\delta\ell_{hf}^2$. Its contribution to the hyperfine 
splitting is $-16\alpha^6\mun^3/3m^2n^4$. Here, however, the NRQED 
results give only half that value, together with other terms, to be discussed 
in the next section.

Before leaving the subject, we wish to propose a minor modification of the 
binding correction to the leptonic $g$-factors that has been used in the 
measurement of the muon magnetic moment in muonium (Liu et al 1999). To order 
$\alpha^2$ and to second order in $m_1/m_2$, this correction has been 
calculated (Grotch and Hegstrom 1971, Faustov 1970) as
\beg{24}g'_i\approx g_i\left[1-\frac{\alpha^2}{3}+\alpha^2\frac{m_1}{2m_2}
\left(1-2\frac{m_1}{m_2}\right)\right],\en
where $g_i$ (i=1,2) are the $g$-factors of the two free leptons. The 
modification consists of replacing the mass factors by $\mun/m$.
More precisely, we propose
\beg{25}g'_i=g_i\frac{2\gamma+1}{3}\,\frac{m}{E},\quad E\approx m -\frac{\mun 
\alpha^2}{2n^2}.\en
The reason is that the Zeeman operator must be an odd function of $E$. In fact,
any small change $\delta E$ caused by $\cal CPT$-invariant perturbations must 
be odd in $E$. In the case at hand, $\cal CPT$-invariance requires equal energy
levels for muonium and antimuonium, even in the presence of a magnetic field. 
Both systems are described by a single eigenvalue equation, with $E^2$ as 
eigenvalue (Malvetti and Pilkuhn 1994). First-order perturbation theory 
produces a small shift $\delta (E^2)$, from which $\delta E$ follows as in eq.
(\ref{13a}). For the hyperfine operator 
$\bld{\sigma}_{hf}^{(1)}$ of equation (\ref{12}), $\delta (E^2)$ goes as 
$E^{-2}$, due to $\mu/E=m_1m_2/E^2$. For the Zeeman operators, $\delta(E^2)$ 
is 
obviously independent of $E$, to order $\alpha^2$. For positronium, (\ref{25})
has been verified by a NRQED calculation (Grotch and Kashuba 1973). 
\end{section}

\begin{section}{The Gupta hyperfine operator}\label{ch4}
According to the NRQED calculation to order $\alpha^6$, the total hyperfine
splitting $\Delta E_{hf}=E$(triplet)-$E$(singlet) contains also the 
combination $-16\alpha^6\mun^3(\ln(\alpha/n)+\Psi +C+7/6-1/2n)/3n^3m^2$.
The last term in the bracket may be decomposed into $-1/n+1/2n$, where the 
$-1/n$  arises from the previously mentioned second-order hyperfine interaction. Comparison with (\ref{4}) shows that the $n$-dependence of the remaining part
is identical with that of the ``Gupta operator''. The complete 
$\alpha^6$-result for S-states is 
\beg{27}\Delta E^{(6)}_{hf,NRQED}=\Delta E^{(6)}_{hf,D}+
\Delta E^{(6)}_{hf,G}+8\alpha^6\mun^3F'_{hf}/3n^3m^2,\en
where $\Delta E^{(6)}_{hf,D}$ contains all terms from the two-fermion
Dirac equation, and
\beg{28}\Delta E^{(6)}_{hf,G}=-\frac{16}{3}\alpha^3\frac{\mun}{m}\langle
p_G^2/2\mu\rangle,\quad F'_{hf}=\frac{91}{36}+\frac{13}{2}\ln\frac{3}{4}
-2\ln 2+f_{hf},\en
 where $f_{hf}$ is a numerical function of $\mun/m$, which is
presently known in analytic form only for $\mun/m=0$ and $1/4$. The ``Gupta''
part of (\ref{28}) is the nonrelativistic expectation value of an operator
that may be combined with the Dirac hyperfine operator in (\ref{12}):
\beg{29}\bld{\sigma}_{hf}=\bld{\sigma}_{hf}^{(1)}[1+2(\mu/E)\alpha^2
\ln(\gamma\rho)].\en
With $\bld{\sigma}_{hf}^{(1)}=
-i\bld{\sigma}_1\times\bld{\sigma}_2\alpha\mu/E\rho$,
the Gupta hyperfine operator goes as $\rho^{-1}\ln\rho$, which is even more
singular than the $\rho^{-1}$ of the Dirac hyperfine operator. However, as it
is only a first-order correction, it may be combined with the Dirac hyperfine
operator into less singular forms, for example
\beg{30}\bld{\sigma}_{hf}=-i\bld{\sigma}_1\times\bld{\sigma}_2Vm_1m_2
[E^2-2m_1m_2\alpha^2\ln(\gamma\rho)]^{-1}.\en
This combination goes as $(\rho\ln\rho)^{-1}$ for $\rho\to 0$, which appears
to admit a nonperturbative use in the Dirac equation. 

This extension is of little importance for QED bound states. But
hyperfine operators appear also in quarkonium models, mainly in the form of
Breit operators for heavy quarkonium. For the vector $(1^-)$ and pseudoscalar
$(0^-)$ mesons, one expects a small and constant hyperfine splitting in $E^2$,
$\Delta=E^2(1^-)-E^2(0^-)$ (Mannel 1998). However, a closer look reveils that 
$\Delta$ increases with decreasing meson masses, from 0.48 GeV$^2$ for the 
heavy b quarkonium (Review of Particle Physics 1998) up to 0.57 GeV$^2\approx 
m_\rho^2$ for the $\rho-\pi$
system, where NRQED expansions would diverge.
Moreover, $m_\rho^2/m_\pi^2=30$ excludes a perturbative treatment of the hyperfine operator. 
It is true that the quark model for pions must differ drastically from any QED analogue, for example in its dependence on $m_1$ and $m_2$.
But the pion is only the lightest meson in a long list of light mesons
whose quantum numbers all agree with the naive quark model. Until the masses
of these mesons are calculated by fundamental methods such as lattice
QCD, it may be allowed to replace the nonrelativistic QCD
Breit operators by the relativistic QCD hyperfine
operator. This operator has a similar structure as in QED. Even in the 
absence of a detailed quarkonium model,
the experimental increase of $\Delta$ hints at an increase of the hyperfine
operator for small $E^2$, in agreement with (\ref{30}).

One of the authors (H.P.) would like to thank Dr. J.O. Eeg for the kind 
hospitality extended to him at the physics institute of the University of 
Oslo, where this work was begun.
Helpful comments by Th. Mannel and K. Melnikov are gratefully acknowledged.
This work has been supported by the Volkswagenstiftung.
\end{section}

\appendix
\begin{section}{Rederivation of the two-fermion equation}
The standard 16-component formalism for two fermions contains two kinetic 
energy operators $\bld{\alpha}_i\bld{p}_i$ (i=1,2), with $\bld{\alpha}_i=
\gamma^5_i\bld{\sigma}_i$ and $\bld{\sigma}_i=$ Pauli spin matrices. The 
essential point of the eight-component formalism in the cms $(\bld{p}_1=-
\bld{p}_2=\bld{p})$ is the absence of $\bld{\alpha}_2\bld{p}_2=-\gamma^5_2
\bld{\sigma}_2\bld{p}$ in the free two-body equation, to which the interaction 
operator is added. The desired equation is easily constructed from the 
kinematical constraints: The direct product $\psi_0=\psi_1\otimes
\psi_2$ of two free particle solutions $\psi_1$ and $\psi_2$
satisfies two Klein-Gordon equations, which are required in the cms, where one 
may use $\bld{p}_1^2=\bld{p}_2^2\equiv \bld{p}^2$:
\beg{a1}(p^{02}_1-m_1^2)\psi_0=(p^{02}_2-m_2^2)\psi_0=\bld{p}^2\psi_0.\en
After elimination of $p_1^0-p_2^0=i\partial_{t1}-i\partial_{t2}$
using $(p_1^0-p_2^0)\psi_0=(m_1^2-m_2^2)(p_1^0+p_2^0)^{-1}\psi_0$,
 there remains
a single equation containing $p^0_1+p^0_2\equiv E$:
\beg{a2}(k^2-\bld{p}^2)\psi_0=0,\quad 4E^2k^2=E^4-2E^2(m_1^2+m_2^2)+(m_1^2-
m_2^2)^2.\en
The equation applies to particles of any spins. For fermions, it must be 
linearized in $\bld{p}$. Here it suffices to write $\bld{p}^2=(\gamma_5
\bld{\sigma}_1\bld{p})^2$, with $\gamma_5^2=1$ and one set of Pauli matrices, 
$\bld{\sigma}_1$. At the same time, $k^2$ must be factorized, writing 
\beg{a3}k^2=(\varepsilon-\mu\beta)(\varepsilon+\mu\beta), \quad \beta^2=1.\en
For two structureless fermions, the symmetry of $k^2$ under the exchange $m_1^2
\leftrightarrow m_2^2$ must be preserved in the factorization, which leads to 
the expressions (\ref{1}) for $\mu$ and $\varepsilon/\mu$. In a last step, the 
$E$ is removed from the denominators of $\mu$ and $\varepsilon$:
\beg{a4}\left(E\varepsilon-E\mu\beta-\gamma_5\bld{\sigma}_1E\bld{p}\right)
\psi_0=0,\quad E\varepsilon=\frac{1}{2}(E^2-m_1^2-m_2^2), \quad E\mu=m_1m_2.\en
The factor $E$ in front of $\bld{p}=-i\nabla$ is absorbed by a rescaling of the
variable $\bld{r}$, after which (\ref{a4}) is an explicit eigenvalue equation 
for $E^2$. A slightly more convenient dimensionless form is obtained by 
dividing by $E\mu=m_1m_2$ as in equation (\ref{5}).

In the 16-component formalism, the direct product $\psi^{(16)}_0= \psi_{1D} 
\otimes \psi_{2D}$ of two free Dirac spinors satisfies two free Dirac 
equations,
\beg{a5}
(p_i^0-\gamma^5_i \bld{\sigma}_i {\bf p}_i - m_i \beta_i )\psi_0^{(16)} = 0,
 \quad \beta_{i} = \gamma_{i}^{0}\, , \quad p_i^{0} =i \partial_{ti}.\en
The sum of these equations in the cms gives, with
$\bld{p}_1=-\bld{p}_2\equiv \bld{p}$,
\beg{a6}
[E-( \gamma_1^5\bld{\sigma}_1-\gamma_2^5\bld{\sigma}_2)\bld{p} - m_1 \beta_1 - 
m_2 \beta_2] \psi^{(16)}_0=0.\en
In the following, (\ref{a6}) is transformed into (\ref{a4}) before the 
corresponding interaction operator is added. The motivation for this step comes
from the details of the interaction, but the main point is easily stated: In 
a first approximation, the differential approach with interaction included 
replaces the $E$ in (\ref{a6}) by $E-V^{(1)}$, where $V^{(1)}=-\alpha/r$ is the
main part of the Fourier transform of the first Born approximation. When this 
form is reduced to (\ref{a4}), then $E^2$ in $E\varepsilon$ is replaced by 
$(E-V^{(1)})^2$. The second-order Born approximation provides 
several additional operators, which in leading order cancel the squares of 
the first-order operators, such as the 
$(V^{(1)})^2$ in $(E-V^{(1)})^2$. These cancellations occur not only in the 
differential equation scetched here, but also in the Bethe-Salpeter equation 
and in NRQED calculations. 
When the interaction is added in the eight-component form (\ref{a4}), $E^2$ is 
automatically replaced by $E^2-2EV^{(1)}$; the operator $(V^{(1)})^2$ is 
absent. In this sense, the first Born approximation in the eight-component
scattering includes the leading terms of the second Born approximation in the 
16-component scattering.\\
To achieve the reduction from (\ref{a6}) to (\ref{a4}), 
$\psi^{(16)}_0$ is divided into two octets $\psi_{0P}$ and $\chi_{0P}$, which
have $\gamma_{1}^{5} = \gamma_{2}^{5} \equiv \gamma_{5}$ and $\gamma_{1}^{5} =
- \gamma_{2}^{5} = \gamma_{5}$, respectively. The round bracket is $\gamma_5
(\bld{\sigma}_1-\bld{\sigma}_2)\equiv\gamma_5\Delta\bld{\sigma}$ when acting on
$\psi_{0P}$, and $\gamma_5
(\bld{\sigma}_1+\bld{\sigma}_2)\equiv\gamma_5\bld{\sigma}$
when acting on $\chi_{0P}$. In the chiral basis,
$\gamma_{1}^{5}$ and $\gamma_{2}^{5}$ are diagonal:
\beg{a7}\gamma_{i}^{5} = \left( \begin{array}{cc}1 & 0 \\
0 & -1 \end{array} \right)_i , \,\,\beta_i =
\left( \begin{array}{cc} 0 & 1 \\
1 & 0 \end{array} \right)_i  , \,\,
\psi_i = \left( \begin{array}{c}\psi_{ir} \\\psi_{il}\end{array} \right) , \,\,
\psi_{0P} = \left( \begin{array}{c}\psi_{rr} \\
\psi_{ll} \end{array} \right) ,  \,\, 
\chi_{0P} = \left( \begin{array}{c}\psi_{rl} \\
\psi_{lr} \end{array} \right)  , \,\,\en
where the indices $r$ and $l$ (= righthanded, lefthanded) refer to the
eigenvalues $\pm 1$ of $\gamma_{i}^{5}$.
Each $\beta_i$ exchanges $ir$ with $il$; $\beta_2$ exchanges $\psi_{0P}$ with 
$\chi_{0P}$, while $\beta\equiv\beta_1\beta_2$ exchanges each upper quartet 
with the lower one.
\beg{a8}\beta_2 \psi_{0P} = \chi_{0P} \, ,\quad \beta_2 \chi_{0P} =
\psi_{0P}\quad\beta\gamma_5+\gamma_5\beta=0.\en
Consequently, (\ref{a6}) is decomposed as follows:
\beg{a9}(E - \gamma_5\bld{p}\Delta\bld{\sigma})\psi_{0P} = m_{+} \chi_{0P},
 \quad (E-\gamma_5\bld{p\sigma}) \chi_{0P} = m_{+} \psi_{0P}, \quad m_\pm=
m_2\pm\beta m_1 .\en
Using the first equation for the elimination of $\chi_{0P}$, one obtains for the second one
\beg{a10}(E- \gamma_5\bld{p}\bld{\sigma}) (m_{+})^{-1}(E-\gamma_5
\bld{p}\Delta\bld{\sigma}) \psi_{0P} =m_{+} \psi_{0P}.\en
Multiplying this equation by $m_{+}$ and using 
\beg{a11} m_{+} \gamma_5 = \gamma_5 m_{-},
\quad (\bld{p\sigma})(\bld{p}\Delta\bld{\sigma})=0,\en
one arrives at the following equation
\beg{a12}[E^2-E\gamma_5(\bld{p\sigma}m_-/m_++\bld{p}\Delta\bld{\sigma})-m_+^2
]\psi_{0P}=0.\en As a rule, an elimination of components produces
operators of second order in $\bld p$. This is prevented here by (\ref{a11}).
The factor $m_-/m_+$ is removed by the transformation
\beg{a13}\psi_{0P} = c \psi_0, \quad c^{-1} \gamma_5 = \gamma_5c, \quad c
\bld{\sigma} c = \bld{\sigma} m_{+} /m_{-}\, , \quad c \Delta \bld{\sigma} c =
\Delta \bld{\sigma}.\en
\begin{equation}\label{a14}
c = (m_{+} m_{-}) ^{-1/2} \left\lbrack m_2 +  m_1 \beta (1 +
\bld{\sigma}_1 \bld{\sigma}_2 )/2 \right\rbrack = ( m_{+} m_{-} ) ^{-1/2} ( m_{+}- 2 m_1 \Lambda_s \beta),\end{equation}
where $\Lambda_s = ( 1 - \bld{\sigma}_1 \bld{\sigma}_2 ) /4$ is the projector
on singlet spin states. The inner bracket in (\ref{a12}) becomes $2\bld{p\sigma}_1$, and division by $2$ produces (\ref{a4}). As mentioned before, the result
is trivial, but the $c$-transformation will be needed again for the $S$-matrix,
$S=1+iT$.
The $16\times 16$ form of the fermion-fermion $T$-matrix in the cms is
$T_{if}=\overline u_1'\overline u_2'\hat Tu_1u_2$, where $u_i$ and 
$\overline u_i'$ are the free Dirac spinors for the in- and outgoing fermions,
$\psi_i=u_ie^{i\phi_i},\;\phi_i=\bld{k}_i\bld{r}_i-E_it$. In the $(\psi,\chi)$
basis, analogous expressions are defined for the ingoing $\psi_{0P}$ and 
$\chi_{0P}$,
\beg{a15}\psi_{0P}=ve^{i\phi},\quad\chi_{0P}=we^{i\phi},
\quad \phi=\phi_1+\phi_2,\en
and for the outgoing $\psi_{0P}'^\dagger,\;\chi_{0P}'^\dagger$. $T_{if}$ is now
expressed as $v'{}^\dagger T_v v,\;v'{}^\dagger T_{vw}w,
\; w'{}^\dagger T_{wv}v$ and $w'{}^\dagger T_w w$. The first Born
approximation $T_{if}^{(1)}$ has $T_{vw}=T_{wv}=0$. The 
product of Dirac matrices $\gamma_1^\mu\gamma_{2\mu}$ is $\beta_1\beta_2(1-
\gamma_1^5\bld{\sigma}_1\gamma_2^5\bld{\sigma}_2)$, with $\gamma_1^5\gamma_2^5=+1$ in the $v$-components and $-1$ in the $w$-components, respectively.
Remembering $\overline u_i'=u_i'^\dagger\beta_i$ and $\beta_1^2=\beta_2^2=1$,
one finds
\beg{a16}T_{if}^{(1)}=-4\pi\alpha \bld{q}^{-2}[v'^\dagger(1-\bld{\sigma}_1\bld{\sigma}_2)v+w'^\dagger(1+\bld{\sigma}_1\bld{\sigma}_2)w],\en
with $\bld{q}=\bld{k}-\bld{k}'\; (q^0=0)$,
where $\bld{k}$ and $\bld{k}'$ are the in- and outgoing momenta of particle 1.
In general, $T_{vw}$ and $T_{wv}$
appear only for an odd number of matrices $\beta_i$.

One-loop graphs contain two fermion propagators, the product of which may be 
written symbolically as
\beg{a17}P=(/\hspace{-0.23cm}p-m)^{-1}X(/\hspace{-0.23cm}p'-m')^{-1}=
(/\hspace{-0.23cm}p+m)X(/\hspace{-0.23cm}p'+m')/(p^2-m^2)(p'^2-m'^2),\en
where $X$ may be any operator. When both propagators occur on one fermion line
$i$ as in radiative corrections, one has $m=m'=m_i$ otherwise $m=m_1,\;m'=m_2$.
The product (\ref{a17}) may be decomposed as follows:
\beg{a18}P=P_{++}-P_{--},\quad P_{++}=2(/\hspace{-0.23cm}pX
/\hspace{-0.23cm}p' +mm'X)/(p^2-m^2)(p'^2-m'^2),\en
\beg{a19}P_{--}=(/\hspace{-0.23cm}p-m)X(/\hspace{-0.23cm}p'-m')/
(p^2-m^2)(p'^2-m'^2)
=(/\hspace{-0.23cm}p+m)^{-1}X(/\hspace{-0.23cm}p'+m')^{-1}\en
For bound states, $P_{--}$ is of higher order in $\alpha$ because it contains
a product of two antifermion poles. The leading radiative corrections contain
no $\beta$, just as (\ref{a16}). Two-photon exchange is linear in $\beta$,
because $/\hspace{-0.23cm}pX/\hspace{-0.23cm}p'$ contains $\beta_1\beta_2=\beta$ in this case.\\
$T_{if}$ can always be put into the form $w'^\dagger Mv$ by means of 
(\ref{a9}). For $P_{--}\approx 0$, one finds
\beg{a20}M=T_wm_+^{-1}(E-\gamma_5\bld{k}\Delta\bld{\sigma})+(E-\gamma_5
\bld{k}'\bld{\sigma})m_+^{-1}T_v.\en
Before insertion into (\ref{a4}), $M$ must also be $c$-transformed.\\
The form $w'^\dagger Mv$ is no longer hermitian. It may be compared with the
$2\times 2$ single-fermion scattering matrix $T_{if}=u_l'^\dagger M_s u_r$,
which is also complete and nonhermitian. 

The first Born approximation (\ref{a16}) gives
\beg{a21}m_+M^{(1)}=-4\pi\alpha\bld{q}^{-2}2(E-i\gamma_5\bld{k}\,\bld{\sigma}_1
\times\bld{\sigma}_2).\en
Notice the absence of $\bld{k}'$. Its Fourier transform produces the 
interaction operators, which in the dimensionless variable $\bld{\rho}=\mu 
\bld{r}$ lead to equation (\ref{12}).
Thus the only addition to the almost trivial Dirac-Coulomb operator (\ref{5})
is a hyperfine operator, which is not totally unexpected either. Apart from
the replacement $m_2^{-1}\to m^{-1}$, which was already found by Breit, it differs from the standard hyperfine operator in two respects: the 
hermitization has been ``forgotten'', and the dimensionless form has
$m^{-2}$ replaced by $E^{-2}$. 
\end{section}


\begin{references}

Bechert K and Meixner J 1935, Ann. Physik (Leipzig) {\bf 22}, 525\\
Braun MA 1973, Sov. Phys. JETP {\bf 37}, 211\\
Brezin E, Itzykson C and Zinn-Justin J 1970, Phys. Rev. D {\bf 1}, 2349\\
Caswell WE and Lepage GP 1986, Phys. Letters B {\bf 167}, 437\\
Czarnecki A, Melnikov K and Yelkhovsky A 1999, Phys. Rev. Letters {\bf 82}, 
311; Phys. Rev. A {\bf 59}, 4316 \\
Erickson GW 1977, J. Phys. Chem. Ref. Date {\bf 6}, 833\\
Faustov R 1970, Phys. Letters B {\bf 33}, 422\\
Grotch H and Hegstrom RA 1971, Phys. Rev. A {\bf 4}, 59\\
Grotch H and Kashuba R 1973, Phys. Rev. A {\bf 7}, 78\\
Gupta SN et al 1989, Phys. Rev. D {\bf 40}, 4100\\
H\"ackl R, Hund V and Pilkuhn H 1998, Phys. Rev. A {\bf 57}, 3268; 1999 Err. A 
{\bf 60}, 725\\
Kinoshita T 1998, preprint hep-ph/9808351\\  
Liu W et al 1999, Phys. Rev. Letters {\bf 82}, 711\\
Malvetti M and Pilkuhn H 1994, Phys. Rep. C {\bf 248}, 1\\
Mannel Th 1998, Acta Physica Polonica B {\bf 29}, 1413\\
Pachucki K 1997, Phys. Rev. A {\bf 56}, 297; Phys. Rev. Letters {\bf 79}, 
4021\\
Pachucki K and SG Karshenboim 1998, Phys. Rev. Letters {\bf 80}, 2101\\
Pilkuhn H 1995, J. Phys. B {\bf 28}, 4421\\
Review of Particle Physics 1998, European Physical Journal C {\bf 3}, 1\\
Rose ME 1961, Relativistic Electron Theory (Wiley)\\
Sapirstein JR and Yennie D 1990, in: Quantum Electrodynamics
(World Scientific, Singapore)

\end{references}
\end{document}